# Near-field radiation assisted smart skin for spacecraft thermal control

Deyu Xu[a], Junming Zhao[a, b,*], Linhua Liu[c]

[a]*School of Energy Science and Engineering, Harbin Institute of Technology, Harbin 150001, China*
[b]*Key Laboratory of Aerospace Thermophysics, Ministry of Industry and Information Technology*, *Harbin 150001, China*
[c]*School of Energy and Power Engineering, Shandong University, Qingdao 266237, China*

## ABSTRACT

Thermal control is of critical importance for normal operation of spacecraft. Given thermal radiation is the only means of heat dissipation in space, an efficient thermal control approach for spacecraft is to coat the radiator with a tunable-emittance "skin" that can tune its heat dissipation according to various thermal conditions. The existing schemes solely relying on far-field thermal radiation, which are based on mechanical, electrochromic or thermochromic working principles, are difficult to combine the advantages of all-solid-state structure, actively and accurate tuning, and large tuning range of heat flux. In this work, we propose a near-field radiation assisted (NFRA) smart skin for thermal control which can tune the heat rejection accurately and in a large range. It contains a metal-insulator-semiconductor (MIS) structure, where the carrier distribution in the semiconductor layer can be electrically altered. In this way, the near-field heat flux, and ultimately the skin emission power expressed using effective emittance, can be controlled as a function of the applied voltage. The variation range of the effective emittance can exceed 0.7 when adjusting the applied voltage from -10 V to 100 V with our preliminary design. This work opens a new way of smart skin design for active spacecraft thermal control.

**Keywords**：near-field radiation; active spacecraft thermal control; smart skin; emittance tuning

## 1 Introduction

Thermal radiation is the only way for spacecraft to release heat in the vacuum space environment. The fluctuant space thermal conditions and dynamic internal thermal loads necessitate efficient thermal control schemes, which are designed to maintain the spacecraft temperature within a proper range, typically near room temperature, to ensure the normal performance of its inner components [1-3]. Traditional thermal control techniques involving heaters, thermostats, heat pipes, mechanical louvers, etc., are considered to be reliable [4, 5]. However, these devices tend to be too massive and power consumptive, hence difficult to scale well to meet the constrained power and mass budgets of micro/nano-spacecraft, which is promising for future space missions [6, 7].

---

* Corresponding author. email address: jmzhao@hit.edu.cn

For micro/nano-spacecrafts, due to their low thermal capacitance, are more susceptible to fluctuations in thermal loads and environment. One efficient solution is to tune the heat rejection into space using a tunable-emittance "skin" coated on the spacecraft radiator, in response to the variation of internal and external thermal conditions [6, 8]. According to working principle, the tunable-emittance approaches under study can be classified into three categories, i.e., mechanical radiators [6, 9-12], thermochromic devices [13-20] and electrochromic devices [2, 3, 5, 21-25]. The mechanical radiators adjust the shape of emitting surface by electrical (louvers) or thermal (smart surfaces) means, thus tuning radiative characteristics of the controlled surfaces. The thermochromic devices resort to thermochromic materials whose dielectric functions vary with temperature, and they therefore only support passive tuning. The electrochromic devices use electrochromic materials whose dielectric functions can be changed through external electric field and therefore support active tuning. The working principles, characteristics and performances of existing tunable-emittance schemes are summarized in **Table 1**. Despite their respective advantages, there is still much room for improvement, especially in the aspects of accurate and large-range tuning. This motivates us to explore new ways of designing thermal control skin.

**Table 1** The working principles, characteristics and performances of existing tunable-emittance thermal control schemes.

| Working principle | Materials/structures | Advantages | Drawbacks | Maximum $\Delta\varepsilon$ in literature |
|---|---|---|---|---|
| Mechanical | Microelectromechanical louvers and thermal switch [6, 9-12] | rapid and active tuning; large-range tunable | mobile parts; high-power-consuming | 0.74 (Exp.) [9] |
| | Smart surfaces [26-28] | all-solid-state; self-adaptive tuning; large-range tunable | only working in passive mode | 0.80 (Theory) [26] |
| Electrochromic | Inorganic material [2, 21, 22] | active tuning; large-range tunable | tunable only between two states ("colored" or "bleached") | 0.71 (Exp.) [22] |
| | Organic material [3, 5, 23-25] | | | 0.51 (Exp.) [24] |
| Thermochromic | Vanadium dioxide (VO$_2$) [13-15] | all-solid-state; self-adaptive tuning | passive mode; too high phase-change temperature (341 K) | 0.49 (Exp.) [13] |
| | La (M1, M2) MnO$_3$ [16-20] | | passive mode; too large phase- | 0.62 (Theory) [19] |

| | | | change temperature range (~200 K) | |
|---|---|---|---|---|
| Near-field radiation assisted smart skin (This work) | MIS structure | all-solid-state; active and accurate tuning; large-range tunable; | fabrication difficulty; conduction leakage | 0.71 (Theory) |

In the last column, the maximum emittance variation ($\Delta\varepsilon$) in literature of these schemes are listed. These data are obtained either by theoretical calculation or by experiment measurement. The proposed scheme of this work is also listed for comparison and its performance will be shown later in **Section 4**. Its maximum $\Delta\varepsilon$ (0.71) is obtained in the scenario: in-cold case, 270 K, $\alpha_S = 0.1$, $\varepsilon = 0.9$. This value can be further improved by using daytime radiative cooling materials, as described in detail in **Section 4.3.**

Summarizing existing schemes, one can realize that all attentions are paid to how to better tune the heat flux emitted straightly from the controlled surfaces, solely via far-field thermal radiation. Thermal radiation heat flux at nanoscale distance (near-field) can exceed far-field radiation by several orders of magnitude due to photon tunneling, as demonstrated by numerous studies, both through theoretical analysis and experimental measurement [29-36]. The significant enhancement of heat exchange in the near-field regime are promising to enlarge the variation range of heat flux when being modulated. Meanwhile, near-field thermal radiation brings more efficient ways to regulate radiative heat transfer [37-50]. These advantages inspire us to design a smart thermal control scheme with the help of near-field thermal radiation.

In this paper, a new principle of smart skin for spacecraft thermal control is proposed, which is based on electric tuning of near-field radiative heat flux. The proposed NFRA smart skin contains a metal-insulator-semiconductor (MIS) structure [41, 51-53], which is in cascade arrangement with a vacuum gap, a matching layer coated on a substrate and an outermost optical solar reflector (OSR) layer. The electrically gating of the MIS structure will result in the tuning of near-field radiation heat flux and finally the far-field radiation heat rejection flux. This paper is organized as follows. Firstly, the working principle and theoretical model of the NFRA smart skin for thermal control of spacecraft are presented. Then the design consideration of the NFRA smart skin is discussed. Finally, thermal control performance and the influences of the OSR on the tuning capability of the NFRA smart skin are analyzed, which will provide guidance for future applications.

## 2 Concept and theoretical model of NFRA smart skin

The working scenario of the NFRA smart skin for thermal control and its structure is depicted in **Fig. 1**(a). The NFRA smart skin mainly contains four functional components in cascade arrangement:

MIS part, nanoscale vacuum gap, matching layer coated on a substrate and an outermost OSR layer. The MIS structure is in close contact with the controlled surface and is applied a gating voltage. Specific selection rules of the materials of above layers will be given later in **Section 3**. It is necessary to acknowledge that, the nanoscale vacuum gap is difficult to fabricate at this stage, and in actual application it should be maintained by posts which would cause conduction leakage. These two factors are potential drawbacks of the NFRA skin (also listed in **Table 1**). However, these drawbacks are not fatal, considering the rapid developing of nanotechnology and that the posts cover a very small proportion (< 0.01%) of the total device surface [54-58].

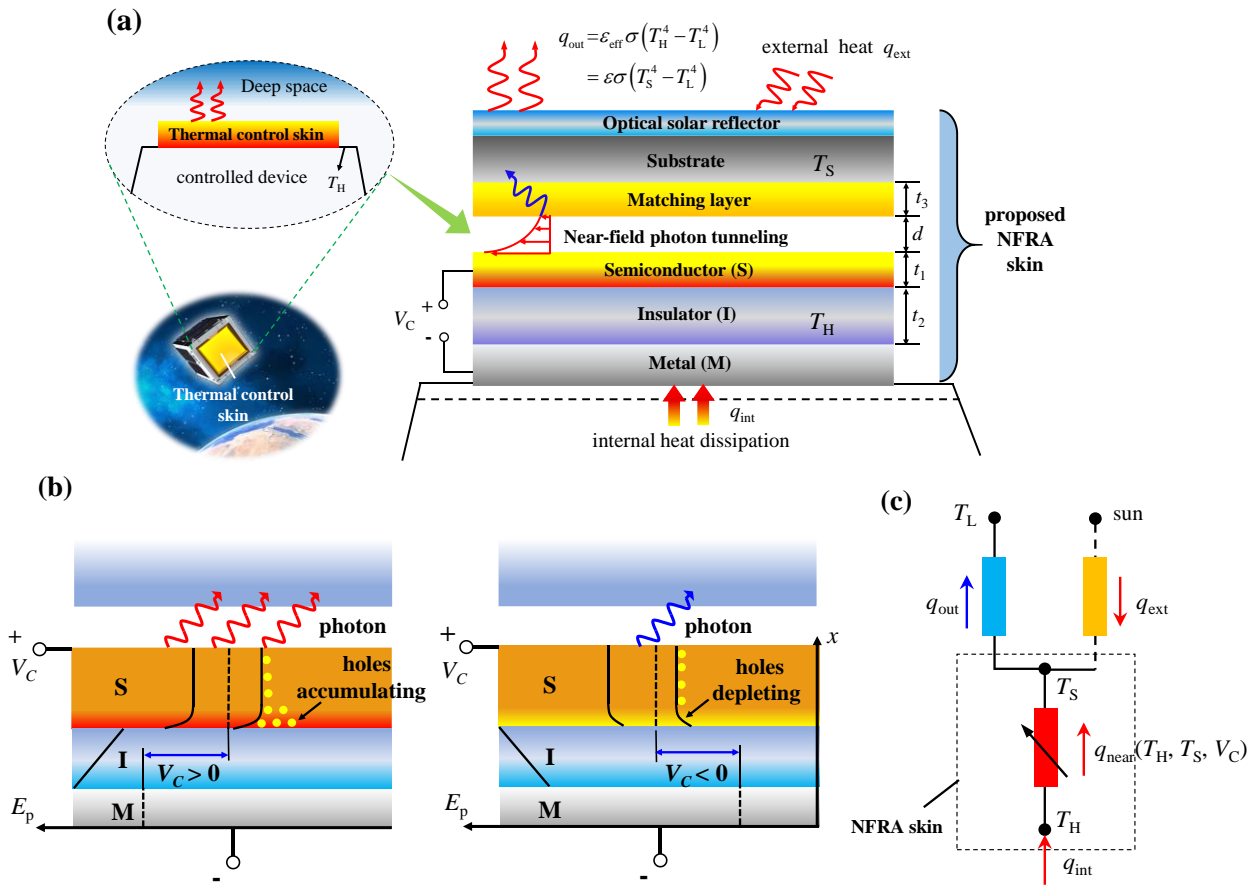

**Fig. 1 (a)** Working scenario of thermal control skin and the structure of the proposed NFRA smart skin. It consists of a gated metal/insulator/semiconductor (MIS) structure, a nanoscale vacuum gap, a matching layer on a substrate and an outermost optical solar reflector (OSR). **(b)** Sketch to show the mechanism of tuning near-field heat transfer using an electric gating. The p-type semiconductor is taken as an example here. **(c)** Thermal resistance network describing the heat transfer of the NFRA smart skin.

## 2.1 Working principle of NFRA smart skin

The working principle of the proposed NFRA thermal control scheme is to tune the near-field thermal resistance between the MIS structure and the matching layer via electric gating, which then results in tuning of heat emission through far field radiation to deep space (heat rejection), due to the cascade arrangement of the thermal resistances between $T_H$ (controlled surface temperature) and $T_L$ (deep space temperature), as shown in **Fig. 1 (c)**.

To better evaluate the performance of the proposed NFRA thermal control skin and compare it with existing devices, an effective emittance ($\varepsilon_{eff}$) is introduced, which measures the ability of heat rejection of the skin. For a given $T_H$, when $V_C$ changes, the near-field radiation resistance changes. Then $T_S$ will change to attain the steady state of heat transfer. Change in $T_S$ certainly leads to the variation of heat rejection to deep space due to Stefan-Boltzmann law. That is, for a given $T_H$ and $V_C$, it is the steady-state $T_S$ that determines the heat rejection. However, in actual application, it is the temperature of the controlled surface $T_H$ that makes sense, instead of $T_S$. The purpose of introduction of effective emittance is to express this heat rejection using $T_H$ (instead of $T_S$). The derivation of $\varepsilon_{eff}$ is described as follows.

When the gating voltage ($V_C$) changes, $T_S$ will adaptively vary and finally the system will attain the steady state of heat transfer, where the heat flux input into and output from the $T_S$ part are balanced (see **Fig. 1 (c)**):

$$q_{near}(T_H, T_S, V_C) + q_{ext} = q_{out} \tag{1}$$

where, $q_{ext}$ represents the absorbed heat from the external heat source. The external heat source mainly includes: the solar irradiation, the earth albedo and the earth irradiation [59]. When the controlled surface faces the sun, it receives the solar irradiation with a power density of solar constant ($S = 1367$ W/m$^2$) [60]. The earth albedo and irradiation are dependent on the position of the spacecraft and are omitted here for simplicity, since they are much smaller than $S$ [26]. When the earth is in the middle of the spacecraft and the sun meanwhile the controlled surface faces away from the earth, the spacecraft can neither receive heat from the sun nor from the earth, in which case it encounters the coldest environment. In this paper, we mainly investigate the above two extreme cases when examining the performance of the proposed skin and designate them as the "in-sun" case and the "in-dark" case, respectively. Note that, $q_{ext}$ equals $\alpha_S S$ in the in-sun case while equals 0 in the in-dark case.

The heat rejected to the deep space ($q_{out}$) is from the OSR surface, which equals [8, 26]

$$q_{out} = \varepsilon \sigma \left( T_S^4 - T_L^4 \right) \tag{2}$$

according to Stefan-Boltzmann law. Here, $\varepsilon$ is the infrared emittance of the OSR and $T_L$ is the space temperature (3 K) [26]. The heat flux determined by Eq. (2) can also be expressed as

$$q_{\text{out}} = \varepsilon_{\text{eff}} \sigma \left( T_{\text{H}}^4 - T_{\text{L}}^4 \right) \tag{3}$$

in terms of the controlled surface temperature which equals that of the MIS, $T_{\text{H}}$, neglecting the contact thermal resistance. It can be immediately derived from Eq. (2) and (3) that:

$$\varepsilon_{\text{eff}} = \frac{\left( T_{\text{S}}^4 - T_{\text{L}}^4 \right)}{\left( T_{\text{H}}^4 - T_{\text{L}}^4 \right)} \varepsilon \tag{4}$$

Mathematically speaking, the closer $T_{\text{S}}$ is to $T_{\text{H}}$, the closer $\varepsilon_{\text{eff}}$ is to 1. Thus, under certain given conditions ($V_{\text{C}}$, $T_{\text{H}}$, $q_{\text{ext}}$, etc.), by iteratively finding $T_{\text{S}}$ that satisfies Eq. (1) at steady state, one can finally obtain the effective emittance $\varepsilon_{\text{eff}}$ of the NFRA skin according to Eq. (4).

The mechanism of tuning near-field heat transfer by adjusting the applied voltage $V_{\text{C}}$ is to alter the carrier distribution in the semiconductor layer, as schematically shown in **Fig. 1** (b). $E_{\text{p}}$ represents the potential energy which equals the potential ($\psi$) multiplied by the unit electronic charge ($q$ = 1.602×10$^{-19}$ As). That is, $E_{\text{p}} = -q\psi$, where "-" indicates that electrons are negatively charged. Let us take p-type semiconductor as an example. If $V_{\text{C}} > 0$, i.e., the semiconductor layer and the metal are connected to the anode and the cathode respectively, the Fermi level in the metal is higher than that in semiconductor and the energy-band in the vicinity of the semiconductor/insulator interface is bent upward (the left panel of **Fig. 1** (b)). In this case, holes, which are the majority carriers in p-type semiconductor, accumulate near the interface due to the Boltzmann's statistic $p = N_{\text{A}} \exp(\frac{-q\psi}{k_{\text{B}}T})$, where $N_{\text{A}}$ is the concentration of acceptors and equals the equilibrium concentration of holes in the semiconductor bulk assuming complete ionization [61]. Meanwhile, the minority carriers, electrons, deplete near the interface. At equilibrium, $n$ and $p$ satisfy the pn product, $np = n_i^2$, with $n_i$ being the intrinsic carrier concentration [62]. If $V_{\text{C}} < 0$, i.e., the semiconductor layer and the metal are connected to the cathode and the anode respectively, the scenario becomes different (the right panel of **Fig. 1** (b)). The energy-band is bent downward and holes deplete near the interface while electrons accumulate, causing a depletion layer there. If the magnitude of the negative $V_{\text{C}}$ gets extremely large, the concentration of electrons will exceed that of holes, forming an inversion layer near the interface. More detailed discussions about MIS can be found in Ref. [61-64]. Changes in carrier distribution lead to modulation of dielectric function of the semiconductor layer and hence tailoring of the near-field heat transfer.

## *2.2 Theoretical model*

In the following, theoretical model to analyze the performance of the NFRA smart skin is presented. To solve the near-field heat flux as a function of $V_C$, the voltage-dependent carrier distribution should be firstly determined. Previous studies involving MIS structure tend to model the region where carrier concentration changes as an effective layer whose thickness is comparable to Debye length, where the concentration of carriers is uniform [41, 52, 65-68]. In fact, however, the carrier concentration exhibits a gradient distribution, especially in the proximity of the interface, as has been analyzed above. In this work, for accuracy, we solve the Poisson equation to obtain the gradient distribution of potential, and further the carrier concentration. The one-dimensional Poisson equation reads [61-64]

$$\frac{d}{dx}\left(\varepsilon_{dc}\frac{d\psi}{dx}\right) = -\rho(x) \tag{5}$$

where, $\varepsilon_{dc}$ is the permittivity constant of the medium of interest, $\rho(x)$ is the charge density (Asm$^{-3}$), composed of immobile ionized donors and acceptors and mobile holes and electrons, that is

$$\rho(x) = q[N_D - N_A + p(x) - n(x)] \tag{6}$$

$N_D$, $N_A$, $p$, $n$ are the concentrations of donors, acceptors, holes and electrons, respectively. In this work, we solve the potential distribution by discretizing Eq. (5) using control volume integration [69]. The boundary conditions depend on the applied voltage $V_C$. The relation between $\rho(x)$ and $\psi(x)$ can be established according to Boltzmann's statics, with which $\psi(x)$ can be solved after several iterations [61-64]. Ultimately, one can obtain the concentration distributions of holes and electrons. More details on the numerical calculation can be seen in **Section 1 of Supplementary Information.**

The gradient distribution of carrier concentration leads to gradient distribution of dielectric function in the semiconductor layer. The dielectric functions of typical semiconductors (e.g., doped silicon and transparent conductive oxide) can be described well by Drude model:

$$\varepsilon_{sem} = \varepsilon_\infty - \frac{\omega_p^2}{\omega^2 + i\omega\Gamma} \tag{7}$$

where $\varepsilon_\infty$ is the high frequency permittivity, $\Gamma$ is the scattering rate, and $\omega_p$ is the plasma frequency which depends on the carrier concentration, following $\omega_p = \sqrt{Ne^2/\varepsilon_0 m^*}$ [51, 66, 70]. We divide the semiconductor layer into several effective layers with each of them featuring a dielectric function depending on the corresponding carrier concentration (Eq. (7)). Then the near-field heat transfer model for multilayer system can be employed to deal with the gradient distribution of dielectric function as long as the number of effective layers is enough, as has been introduced in our previous work [71]. The verification of the independence of the number of layers is given in **Section 2 of Supplementary Information.**

The near-field heat transfer flux between the MIS (with several effective layers in the S layer) at $T_H$ and the opposite part at $T_S$ can be calculated by fluctuational electrodynamics applied to multilayer system [72]:

$$q_{near} = \frac{1}{\pi^2} \int_0^\infty d\omega \left[ \Theta(\omega, T_H) - \Theta(\omega, T_S) \right] \int_0^\infty s(\omega, \beta) d\beta \tag{8}$$

where $\Theta(\omega, T) = \hbar\omega / [\exp(\hbar\omega / k_B T) - 1]$ is the mean energy of a Planck oscillator at the angular frequency $\omega$ and the equilibrium temperature $T$, $\hbar$ is the Planck constant divided by $2\pi$, $k_B$ is the Boltzmann constant, $\beta$ is the component of the wavevector parallel to the interface. $s(\omega, \beta)$ is an exchange function composed of the contributions from both s-polarization and p-polarization. In different regions of $\beta$, $s(\omega, \beta)$ takes different forms [33]:

$$s(\omega, \beta) = \begin{cases} \dfrac{\beta(1-\rho_{01}^s)(1-\rho_{02}^s)}{4\left|1 - R_{01}^s R_{02}^s e^{i2\gamma_0 d}\right|^2} + \dfrac{\beta(1-\rho_{01}^p)(1-\rho_{02}^p)}{4\left|1 - R_{01}^p R_{02}^p e^{i2\gamma_0 d}\right|^2}, & (\beta < \omega/c) \\[2mm] \dfrac{\beta \operatorname{Im}(R_{01}^s) \operatorname{Im}(R_{02}^s) e^{-2\operatorname{Im}(\gamma_0)d}}{\left|1 - R_{01}^s R_{02}^s e^{-2\operatorname{Im}(\gamma_0)d}\right|^2} + \dfrac{\beta \operatorname{Im}(R_{01}^p) \operatorname{Im}(R_{02}^p) e^{-2\operatorname{Im}(\gamma_0)d}}{\left|1 - R_{01}^p R_{02}^p e^{-2\operatorname{Im}(\gamma_0)d}\right|^2}, & (\beta > \omega/c) \end{cases} \tag{9}$$

In Eq. (9), Im() takes the imaginary part of a complex number. $\gamma_0 = \sqrt{k_0^2 - \beta^2}$ is the component of the wavevector perpendicular to the interface in vacuum, with $k_0 = \omega/c$ being the vacuum wavevector magnitude. $R_{0i}^j$ is the effective reflection coefficient for $j$-polarization ($j$ = s, p) between vacuum and medium $i$. In this paper, $R_{0i}^j$ is calculated by transfer matrix method (TMM) for multilayer media [72, 73]. $\rho_{0i}^j = \left|R_{0i}^j\right|^2$ is the effective reflectivity.

## 3 Design consideration and implementation example

Here, we focus on the design rules for the functional layers in **Fig. 1**(a) and give an implementation example of the proposed scheme. The distance between MIS and the matching layer is denoted as $d$. The thicknesses of the semiconductor (S) layer, the insulator layer (I) and the matching layer are $t_1$, $t_2$ and $t_3$, respectively. The goal of our considerations is to amplify the regulation effect of voltage on near-field heat flux and further the effective emittance.

### 3.1 Semiconductor layer and matching layer

As for the semiconductor layer (S), many materials such as doped silicon and transparent conductive oxides can be alternatives. For generality, we choose p-type doped silicon, which has been sufficiently studied and widely used in the semiconductor industry for our preliminary design. The Drude model is employed to model the dielectric function of p-doped Si, considering the dependence of mobility on carrier concentration following Ref. [70] and the influence of temperature following Ref. [74]. The semiconductor/insulator interface will introduce electron-boundary scattering which alters the scattering rate in Drude model of Si film. For metal (such as Au), the additional scattering rate ($\Gamma_{eb}$) can be determined via model fitting with the transient thermo reflectance (TTR) measurement data [75, 76]. It is even difficult to determine $\Gamma_{eb}$ of doped Si film with gradient distribution of carrier density used in this work. Therefore, we use the optical properties of bulk, following the treatment of Ref. [41] (also involving optical constant of doped Si film) to estimate the performance of the NFRA skin, and prove its feasibility preliminarily. The intrinsic carrier concentration as a function of temperature of doped Si is obtained from Ref. [61]. We set the initial doping concentration, namely the concentration of acceptors, to $1\times10^{16}$ cm$^{-3}$. It is also the equilibrium carrier concentration in the doped Si bulk assuming that the acceptors ionize completely. This assumption is reasonable due to the large ionization degree of acceptors, according to Ref. [70]. $t_1$ should be sufficiently small to ensure the significant effect of the carrier concentration change on the near-field heat transfer. We therefore choose $t_1 = 10$ nm throughout this work.

The matching layer cooperates with the semiconductor layer, absorbing different tunneling heat flux when $V_C$ is changed. We use p-doped silicon with doping concentration of $1\times10^{20}$ cm$^{-3}$ as the matching layer. The matching layer and the semiconductor layer cannot match well when the applied voltage is zero (or negative, i.e., in depletion mode) due to the significant difference between their carrier concentrations. When $V_C$ gets larger (in accumulation mode), the carrier concentration in the semiconductor layer increases, making it match better with the matching layer, and thus a larger heat flux can be yielded. The thickness of the matching layer, $t_2$, is set to 50 nm, which does not affect the performance essentially.

### 3.2 Insulator layer

The insulator layer (I) should not support surface phonon polaritons (SPhPs), otherwise it may dominate the near-field heat transfer and submerge the surface plasmon polaritons (SPPs) supported by the semiconductor and the matching layer, weakening the tuning effect [77]. There are many materials can be used, such as non-crystalline SiO, non-crystalline $Si_3N_4$, vitreous $As_2Se_3$, diamond and so on, whose real parts of dielectric functions are always positive in the infrared region [78, 79]. On the other hand, the breakdown field strength of the insulator layer limits the maximum voltage that can be applied. To highlight the physical nature and to clarify the influence of $V_C$ on $\varepsilon_{eff}$ without losing generality, we choose an ideal insulator with constant dielectric function of 4 (close to that of $Si_3N_4$) and put aside the possibility of breakdown for the time being. The thickness $t_2$ is set to 100 nm throughout this work.

### *3.3 Metal layer and others*

The metal layer (M) serves as electrode and aluminum is a conventional choice in MIS because it is easy to evaporate and adheres strongly to the insulator [62]. It has little impact on the near-field heat transfer due to the mismatch of its SPP modes with the Planck's distribution and relatively large thickness of the insulator layer. The substrate of the matching layer should not play a significant role in the near-field heat transfer. Therefore, metals such as aluminum, silver and gold can be used, since their SPPs resonance cannot be thermally excited near room temperature [32, 77, 80]. It should be sufficiently thick such that the OSR layer does not participate in the near-field heat transfer.

The OSR is a kind of structure which has been widely used in spacecraft thermal control, featuring low $\alpha_s/\varepsilon$ [1, 4, 81]. Coincidently, recent advances in daytime radiative cooling technology offer us several alternatives capable of serving as the OSR layer, whose emittance can exceed 0.9 with solar absorptance close to zero [82-86]. In this work, we firstly set $\varepsilon= 0.9$ and $\alpha_s= 0.1$ to evaluate the performance of our scheme conservatively. In **Section 4.3**, we will study the influence of the OSR on the emittance variation, as well as showing the performance of the proposed skin with some state-of-art daytime radiative cooling materials used as the OSR layer.

## 4 Thermal control performance of NFRA smart skin

In this section, we will investigate the thermal control performance of the NFRA smart skin designed in **Section 3**. For simulation purpose, the gap distance $d$ must be chosen to calculate the near-field heat flux $q_{near}$. We study the influence of gap distance on the variation of effective emittance ($\Delta\varepsilon_{eff} = \varepsilon_{eff,\ max} -\varepsilon_{eff,\ min}$) adopting the materials and parameters discussed in **Section 3**. The results are shown in **Section 3 of Supplementary Information**. We find that the optimal gap distance which leads to the maximum $\Delta\varepsilon_{eff}$ is at ~13nm. However, it is difficult to achieve such a small gap distance in engineering practice. Therefore, considering the feasibility in practical application without sacrificing the performance seriously, we set the gap distance *d* to 30 nm, which is, to our knowledge, the smallest size realized by plate-plate near-field heat transfer experiments, when conducting following simulations [54]. In real-world applications, one can assemble arrays of small NFRA skins into the required area to overcome the challenge of fabricating large-area device with such a nanoscale vacuum gap, like the design in Refs. [54, 56].

### *4.1 Performance of emittance tuning*

The introduction of $\varepsilon_{\text{eff}}$ in **Section 2.2** allows us to characterize the performance of NFRA skin in terms of emittance, making it convenient to compare the NFRA skin with traditional thermal control schemes. We first focus on the variation of $\varepsilon_{\text{eff}}$ as a function of $V_C$ at different temperatures, in both the in-sun and the in-dark cases.

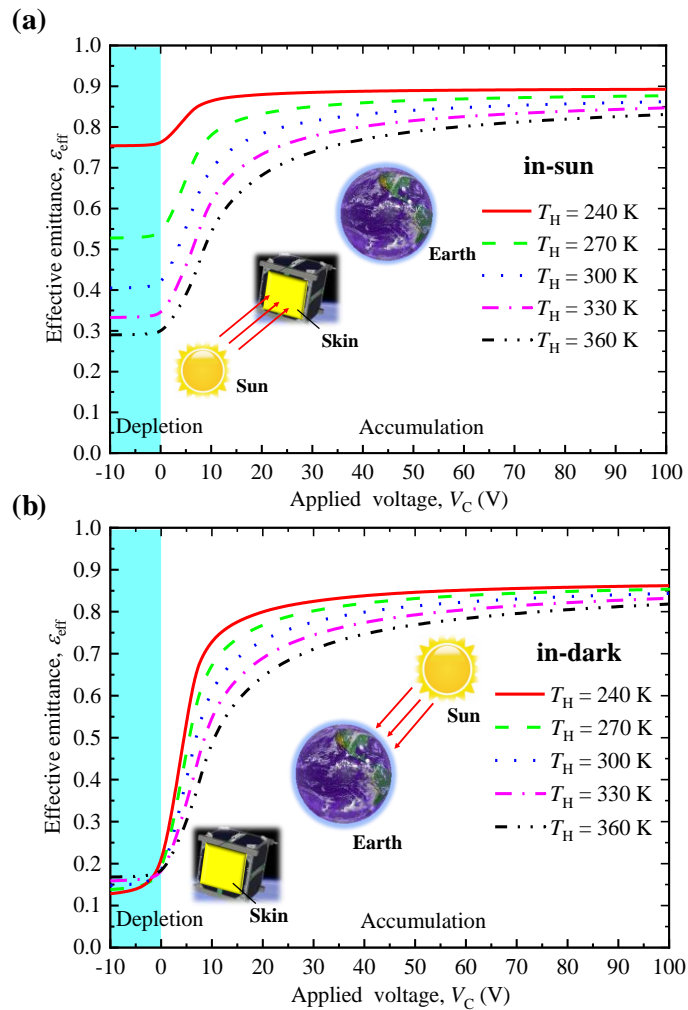

**Fig. 2** Variations of $\varepsilon_{\text{eff}}$ as a function of $V_C$ at different temperatures in the **(a)** in-sun and **(b)** in-dark cases. Five scenarios of $T_H$ = 240K, 270 K, 300 K, 330 K, 360 K are investigated.

Shown in **Fig. 2** (a) and (b) are the variations of $\varepsilon_{eff}$ as a function of $V_C$ at different temperatures in the in-sun and the in-dark cases, respectively. The range of $V_C$ is set to [-10V, 100V], at which the MIS works in accumulation mode ($V_C > 0$) or depletion mode ($V_C < 0$). Too large negative $V_C$ may bring the MIS to inversion mode, which is not considered in this work, since it affects the near-field heat transfer in the same way as accumulation mode, with the roles of electrons and holes inverted [41]. Five scenarios of $T_H$ = 240 K, 270 K, 300 K, 330 K, 360 K are investigated here based on the common operation temperature ranges in spacecraft [1]. The cyan shaded area represents depletion mode, where $\varepsilon_{eff}$ changes little compared to accumulation mode. It can be seen that $\varepsilon_{eff}$ changes continuously with $V_C$. In other words, an arbitrary given $V_C$ uniquely determines an $\varepsilon_{eff}$. This makes the proposed skin stand out in terms of accurate tuning, especially as compared to electrochromic devices where there are only two stable states can be chosen: colored and bleached.

In both cases, $\varepsilon_{eff}$ increases sharply with increasing $V_C$ when $V_C$ is relatively small (< ~20 V) but higher than zero, whereas increases slowly as $V_C$ gets larger. At 300 K, when operating in the range of 0~20 V, $\Delta\varepsilon_{eff}$ can reach over 80 % of that obtained when operating in the range of -10 ~ 100 V. This provides us with an inspiration that the proposed skin can operate in a low voltage range to save power consumption and to avoid breakdown of the insulator, if the required $\varepsilon_{eff}$ is not too large. On the other hand, we note that the variation range of $\varepsilon_{eff}$ is narrowed as the temperature decreases in the in-sun case, due to the drastic increase of $\varepsilon_{eff, min}$ under small $V_C$. Typically, $\varepsilon_{eff, min}$ reaches up to 0.75 at 240 K, reducing $\Delta\varepsilon_{eff}$ to 0.15. This implies that too low temperature may invalidate the NFRA skin. Note that most of devices used in spacecraft work at temperatures higher than 270 K, at which the availability of the proposed skin can be ensured [1, 59] . However, the performance in the in-dark case is not so much dependent on the temperature. Generally speaking, the proposed skin shows better performance in the in-dark case, where $\varepsilon_{eff}$ can vary from 0.14 to 0.85 at 270 K, with $\Delta\varepsilon_{eff}$ = 0.71. Even at 360 K, $\Delta\varepsilon_{eff}$ can still attain as large as 0.65.

To reveal the mechanism responsible for such high performance, we inspect the near-field heat transfer under different $V_C$. The spectral near-field heat flux under $V_C$ = -10 V, $V_C$ = 50 V and $V_C$ = 100 V are shown in **Fig. 3** (a). The temperatures of the emitter and the receiver are set to 310 K and 290 K, respectively. When $V_C$ increases, the heat flux peak moves to lower frequency region and is blue-shifted with increasing $V_C$. Meanwhile, the magnitude of the peak becomes larger as $V_C$ increases. The physical mechanisms for such changes are analyzed in the following paragraphs. In general, when $V_C$ increases, the near-field radiative flux increases, and therefore the resistance due to thermal radiation decreases, and thus the effective emittance increases.

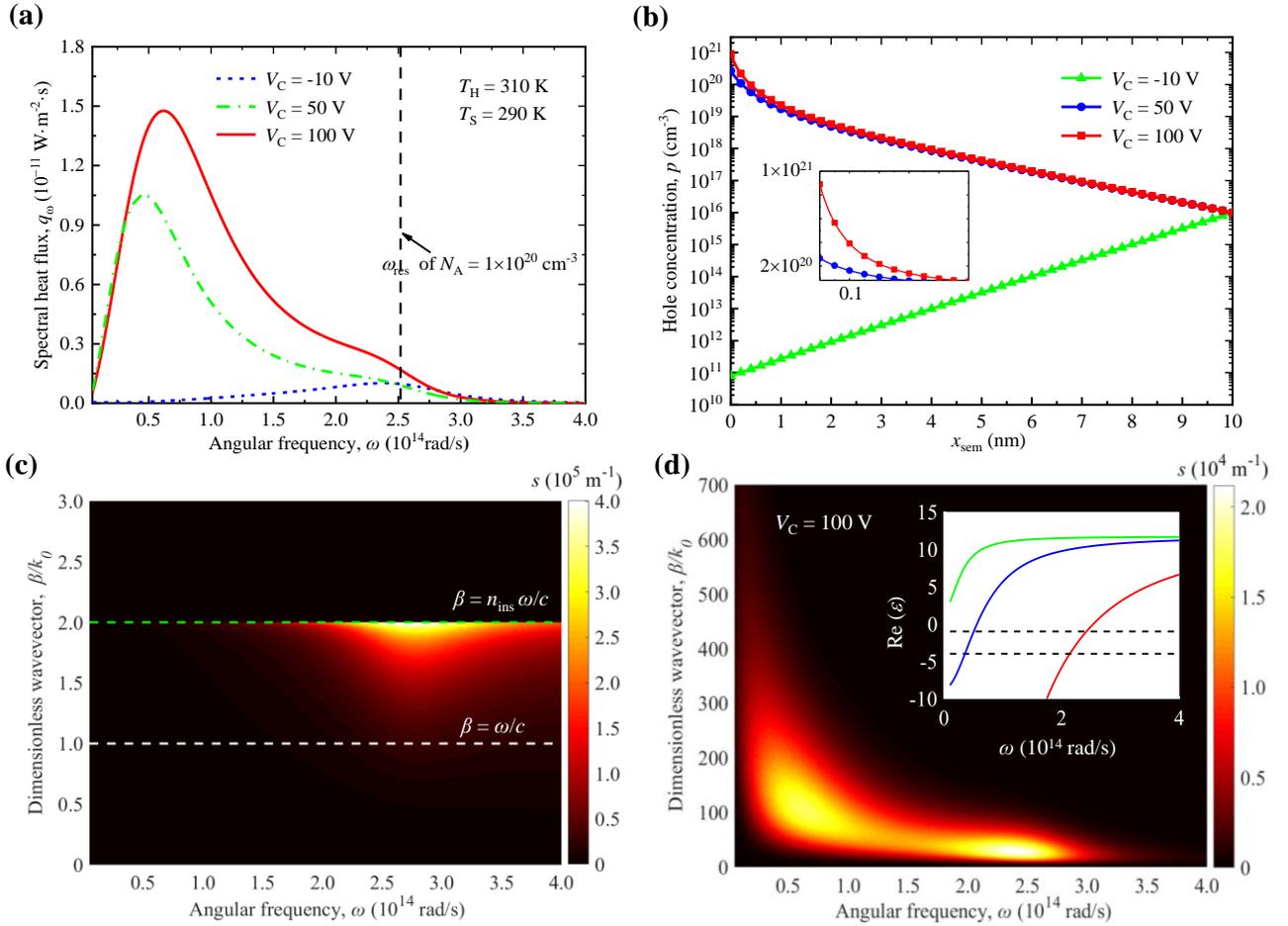

**Fig. 3 (a)** Near-field heat flux spectrum under $V_C$ = -10 V, $V_C$ = 50 V and $V_C$ = 100 V. The temperatures of the emitter and the receiver are 310 K and 290 K, respectively. **(b)** Hole concentration distribution in the semiconductor layer (S) under $V_C$ = -10V, 50 V and 100 V, where, $x_{sem}$ represents the location with respect to the semiconductor/insulator interface. **(c)** Exchange function with respect to $\omega$ and $\beta$ when $V_C$ = -10 V. **(d)** Exchange function with respect to $\omega$ and $\beta$ when $V_C$ = 100 V.

A more thorough understanding of the physical origins requires investigations on the carrier concentration distribution as well as the exchange function given by Eq. (9). In **Fig. 3(b)**, we show the hole concentration distribution in the semiconductor layer (S) under $V_C$ = -10V, 50 V and 100 V. $x_{sem}$ represents the location with respect to the semiconductor/insulator interface. The enlarged inset is also plotted to show the hole concentration in the extreme proximity of the interface. The carrier concentration increases by several orders of magnitude when a positive $V_C$ is applied, attaining $2.7 \times 10^{20}$ cm$^{-3}$ under $V_C$ = 50 V and $8.9 \times 10^{20}$ cm$^{-3}$ under $V_C$ = 100 V, at the semiconductor/insulator interface. Getting far away from the interface, the hole concentration decreases sharply in the range of several nanometers and reaches $N_A$ at the Ohmic contact boundary (see Figure S1 of **Supplementary Information**). Under negative $V_C$, the hole concentration decreases to a very low level (~$7.9 \times 10^{10}$ cm$^{-3}$) near the interface, compared to the initial hole concentration without $V_C$, $N_A$.

**Fig. 3**(c) shows the exchange function determined by Eq. (9) under $V_C$ = -10 V as a function of $\omega$ and $\beta$. The lateral wavevector $\beta$ is normalized by the wavevector in vacuum $k_0$ (= $\omega/c$). A brighter region indicates a larger value of exchange function. The white dashed line and the green dashed line represent the light lines in vacuum ($\beta = \omega/c$) and in the insulator ($\beta = n_{ins}\omega/c$), respectively. In this situation, the semiconductor layer (S) does not support SPPs at its interface between vacuum due to its low carrier concentration. Instead, it is the waveguide modes whose wavevectors are limited by the refractive index of the insulator ($n_{ins}$) that dominate the near-field heat transfer [87]. One can also see that a larger number of electromagnetic states gather in the range of $\omega$ = 2.0~3.5×10$^{14}$ rad/s, compared to other frequency regions. This can be attributed to SPPs at the interface between the matching layer and vacuum, whose resonance frequency is at about 2.50×10$^{14}$ rad/s (obtained by solving Re [$\varepsilon(\omega_{res})$] +1= 0). Although excited at the matching layer/vacuum interface, SPPs with $\beta > n_{ins}\omega/c$ cannot be absorbed by the MIS structure, given the MIS's inability to support surface waves. Thus, only surface modes with $\beta < n_{ins}\omega/c$ can tunnel the vacuum gap and contribute to near-field heat transfer, via playing the role of waveguide modes in the insulator.

However, it becomes a different scenario when $V_C > 0$. Accumulation of holes makes the dielectric characteristic of doped Si ('S' layer) closer to that of metal, with the free carriers coupling strong with electromagnetic waves, i.e., SPPs are supported, both at the interface between vacuum and doped Si and the interface between the insulator and doped Si. Shown in the inset of **Fig. 3**(d) are the real parts of dielectric functions of p-doped Si of different carrier concentration. The upper, middle and lower curves correspond to the carrier concentration of 1×10$^{18}$ cm$^{-3}$, 1×10$^{19}$ cm$^{-3}$ and 1×10$^{20}$ cm$^{-3}$, respectively. The upper and lower dashed straight lines represent Re($\varepsilon$)+1= 0 and Re($\varepsilon$)+$\varepsilon_{ins}$= 0, respectively. The intersections of the curves and the straight lines denote the resonance of SPP at the interfaces between doped Si of corresponding carrier concentration and the corresponding medium (the upper line for vacuum and the lower line for the insulator). With the decrease of carrier concentration, the SPP resonances at both interfaces is red-shifted. Too low carrier concentration makes SPPs unable to be thermally excited in the interested frequency range, as the upper curve indicates.

The gradient distribution of dielectric function in the semiconductor layer give birth to more surface modes with low frequency, through the coupling between the modes supported at the interfaces of layers of different carrier concentrations [71, 88]. Therefore, compared to the case of $V_C$ = -10 V as shown in **Fig. 3**(c), a large bright region of high exchange function can be observed in **Fig. 3**(d) in the frequency region (1×10$^{13}$ rad/s ~ 2×10$^{14}$ rad/s), with wavevectors reaching as large as 700 $k_0$ below $\omega$ = 0.5×10$^{14}$ rad/s. A region of extremely high exchange function spinning the $\beta$ range of 100 $k_0$ ~ 200 $k_0$ can be observed at $\omega = 0.6 \times 10^{13}$ rad/s, which associates with the peak of spectral heat flux in **Fig. 3(a)**. Although the modes near $\omega = 2.5 \times 10^{14}$ rad/s are also of large exchange functions, they, subjected to their small wavevectors, cannot significantly boost near-field heat transfer. This is why the heat flux spectrum no longer peaks near $\omega = 2.5 \times 10^{14}$ rad/s under high applied voltage.

## *4.2 Capability of controlling temperature*

Since the physical mechanism has been clarified, we further study the capability of controlling temperature of the NFRA smart skin. The internal heat generation in a spacecraft mainly originates from propulsion, electronic components, solar cells, etc., which will be finally transferred to the surface for rejection [3, 4, 11, 15, 26]. We denote this internal heat flux with $q_{\text{int}}$. The working conditions of the internal components are always changing, resulting in the fluctuation of internal heat dissipation. To ensure the ability of the surface to reject or preserve heat in all internal thermal conditions, the applied $V_C$ should be adjusted to maintain the surface temperature depending on the variation of $q_{\text{int}}$. At steady state, the input and output heat of the whole skin are equal (see **Fig. 1(c)**), i.e.,

$$q_{\text{int}} + q_{\text{ext}} = q_{\text{out}} \tag{10}$$

$q_{\text{out}}$ is determined by Eq. (2) or (3). The relation between $q_{\text{int}}$ and $V_C$ at different temperatures are shown in **Fig. 4**.

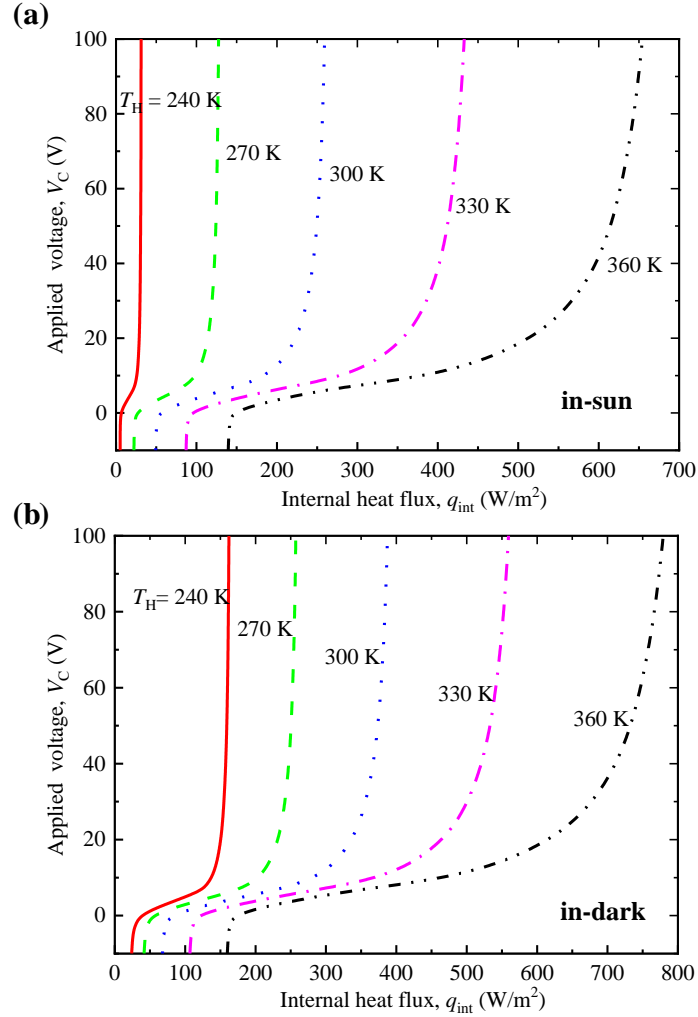

**Fig. 4** Relation between $q_{int}$ and the required $V_C$ to maintain the surface temperature $T_H$ at 240 K, 270 K, 300 K, 330 K, 360 K, in the **(a)** in-sun case and **(b)** in-dark case. See **Fig. 2** for the visual interpretations of the in-sun case and the in-dark case.

When a given $q_{int}$ is input, a specific $V_C$ is required to maintain the temperature at what the curve represents. A higher or a lower $V_C$ relative to the vertical coordinate of one curve would shift the temperature to a lower or a higher one, causing the thermal condition undesired. In both the (a) in-sun and (b) in-dark cases, a larger $q_{int}$ requires a higher $V_C$. This can be easily understood given the positive correlation between $\varepsilon_{eff}$ and $V_C$ as depicted in **Fig. 1**. Similar to the discussion about $\varepsilon_{eff}$, a relatively small variation range of $V_C$ (0~20 V) can cope with the fluctuation of $q_{int}$ over a wide range. From another perspective, when $V_C$ exceeds ~20 V, a small increment of $q_{int}$ requires a significant increase in $V_C$, to keep $T_H$ constant. One can also infer that there exists a maximum allowable $q_{int}$ for a given temperature, above which $T_H$ would increase, no matter how high $V_C$ is. This maximum allowable $q_{int}$ is given by $\varepsilon\sigma\left(T_H^4 - T_L^4\right)$ for a given $T_H$, since $\varepsilon$ is the maximum achievable effective emittance.

Comparing **Fig. 4**(a) and (b), one can find that the in-dark case allows larger $q_{int}$ to enter, under a given $V_C$. This is because in the in-dark case the skin does not absorb the radiative heat flux from the sun, which is present in the in-sun case.

Then we explore the ability of the proposed skin to control temperature in different internal thermal conditions, namely, $q_{int}$. The steady-state temperature of the controlled surface ($T_H$) can be accurately modulated through the tuning of applied voltage, $V_C$, as depicted in **Fig. 5**.

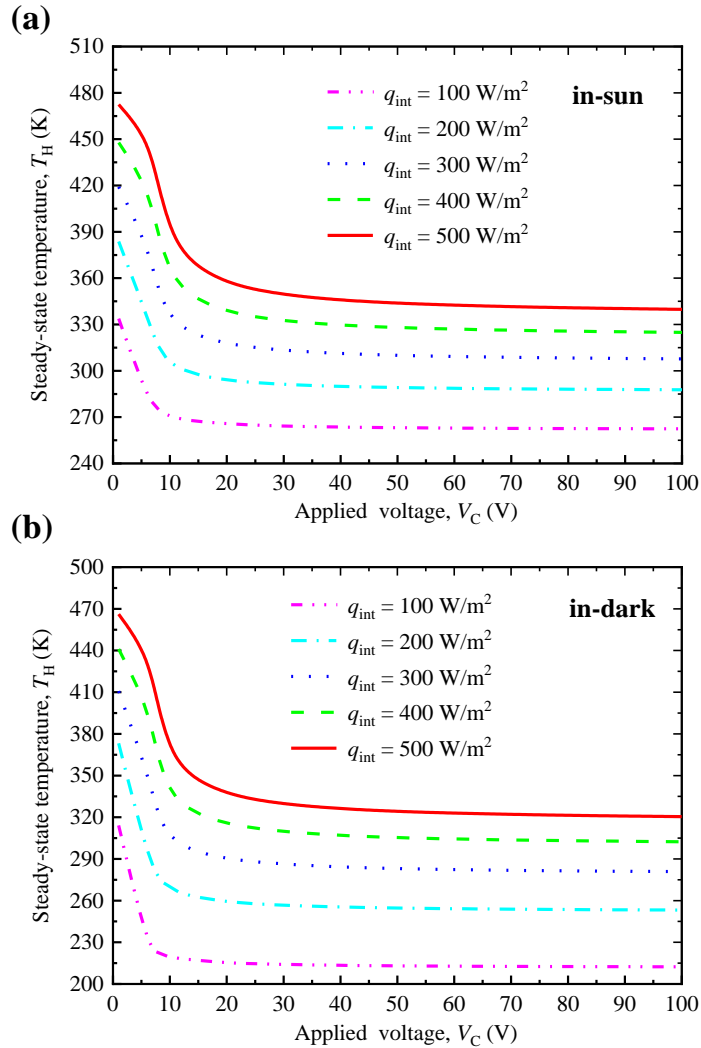

**Fig. 5** Steady-state temperature of the controlled surface ($T_H$) controlled by adjusting $V_C$ from 0 to 100 V (only in accumulation mode), in the **(a)** in-sun case and **(b)** in-dark case. Scenarios of $q_{int}$ = 100 W/m², 200 W/m², 300 W/m², 400 W/m², 500 W/m² are shown in both cases. See **Fig. 2** for the visual interpretations of the in-sun and the in-dark cases.

Here, operations in depletion mode ($V_C$ < 0) are not considered, since they contribute little to the variation of $\varepsilon_{eff}$, as has been demonstrated in **Fig. 2**. We investigate $q_{int}$ = 100 W/m² ~ 500 W/m² in both the (a) in-sun case and (b) in-dark case. Note that, when $q_{int}$ is zero, $T_S$ would increase to $T_H$ to ensure that the MIS does not reject net heat flux to the matching layer at steady state. As a result, $\varepsilon_{\text{eff}} = \varepsilon$, and the equation $\varepsilon\sigma\left(T_H^4 - T_L^4\right) = q_{\text{ext}}$ has to be satisfied. In the in-dark case, $q_{\text{ext}} = 0$ leads to $T_H = T_L = 3$ K, while, in the in-sun case, $q_{\text{ext}} = \alpha_S S$ leads to $T_H = 227.50$ K. That is to say, if $q_{\text{int}}$ is null, adjusting $V_C$ has no effect on the steady-state surface temperature. In practice, however, all the devices working in spacecraft generate heat which needs to be dissipated, i.e., $q_{\text{int}} > 0$, thus allowing the proposed thermal control scheme to control $T_H$ trough $V_C$.

According to the results shown in **Fig. 5**, larger $q_{\text{int}}$ results in higher $T_H$ under a fixed $V_C$, which is intuitive, considering that surfaces with higher temperature radiates more heat to deep space which should be equal to $q_{\text{int}}$ to approach steady state. When $V_C$ is low, typically lower than 10V, $T_H$ decreases significantly with increasing $V_C$, indicating a high tuning performance. It can be found again from **Fig. 5**, that the steady-state $T_H$ is insensitive to $V_C$ higher than 20V. In fact, we can expect the lowest temperature the surface can reach (by increasing $V_C$ to infinity) for a given $q_{\text{int}}$, by solving $\varepsilon\sigma\left(T_{H,\text{min}}^4 - T_L^4\right) = q_{\text{int}} + q_{\text{ext}}$, which yields $T_{H,\text{min}} = \sqrt[4]{\dfrac{q_{\text{int}} + q_{\text{ext}}}{\varepsilon\sigma} + T_L^4}$. $T_{H,\text{min}}$ in the in-sun case, for its non-zero $q_{\text{ext}}$, is higher than that in the in-dark case. From **Fig. 5**, it is known clearly that the temperature range within which $T_H$ can be controlled via electric gating, as well as how to control $T_H$ to a desired value with a known $q_{\text{int}}$, both of which are of significance for spacecraft thermal control.

## *4.3 Influence of OSR on emittance tuning*

Finally, we study the influences of the solar absorptance ($\alpha_S$) and infrared emittance ($\varepsilon$) of the OSR on the performance of the NFRA smart skin. According to its working principle, the effective emittance $\varepsilon_{\text{eff}}$ is tuned through tailoring the near-field heat transfer by $V_C$. As can be inferred from Eq. (1), if $q_{\text{ext}}$ is comparable to or larger than $\varepsilon\sigma\left(T_S^4 - T_L^4\right)$, $q_{\text{near}}$ will lose its contribution to heat transfer, causing the tuning effect invalid. Thus, the solar absorptance $\alpha_S$, which determines the absorbed heat flux from the sun, should not be too high compared with the emittance $\varepsilon$. Here we take the in-sun case, where the impact of the sun is maximized, as an example, then the condition below should be satisfied for a $T_H$ of interest:

$$q_{\text{ext}} = \alpha_S S \leq \varepsilon\sigma\left(T_S^4 - T_L^4\right) \leq \varepsilon\sigma\left(T_H^4 - T_L^4\right) \Rightarrow \frac{\alpha_S}{\varepsilon} \leq \frac{\sigma\left(T_H^4 - T_L^4\right)}{S} \tag{11}$$

For $T_H = 300$ K, it gives the maximum $\alpha_S/\varepsilon$ of 0.336. The variation of $\varepsilon_{\text{eff}}$ ($\Delta\varepsilon_{\text{eff}}$) is shown as a function of $\varepsilon$ and $\alpha_S/\varepsilon$ in **Fig. 6**. When determining $\Delta\varepsilon_{\text{eff}}$, $V_C$ is adjusted from -10 V to 100 V. $\Delta\varepsilon_{\text{eff}}$ can reach as large as 0.78 when $\varepsilon = 1$, $\alpha_S = 0$, which decreases with the decrease of $\varepsilon$ and the increase of $\alpha_S/\varepsilon$. Too low $\varepsilon$ results in a near-zero $\Delta\varepsilon_{eff}$, even though $\alpha_S/\varepsilon$ is zero. $\varepsilon$ larger than 0.67 and $\alpha_S/\varepsilon$ lower than 0.12 should be chosen if one needs a $\Delta\varepsilon_{eff}$ larger than 0.5. The isolines of $\Delta\varepsilon_{eff}$ are approximately concentric ellipses, with the center being (1, 0). That illustrates that, to obtain a specific $\Delta\varepsilon_{eff}$, a lower $\alpha_S$ should be used if $\varepsilon$ is not high enough, vice versa.

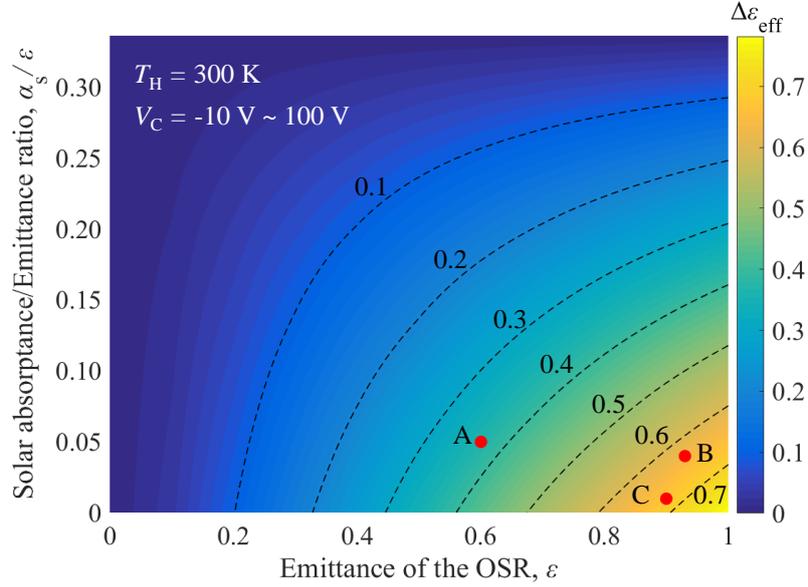

**Fig. 6** Variation of $\varepsilon_{eff}$ ($\Delta\varepsilon_{eff}$) as a function of the OSR emittance $\varepsilon$ and its solar absorptance/emittance ratio ($\alpha_S/\varepsilon$). $T_H$ = 300 K and $V_C$ varies from -10 V to 100 V.

Daytime radiative cooling materials are famous for high infrared emittance and low solar absorptance, thus being ideal choices for the OSR layer [82-86]. Here, we take some typical reported daytime radiative cooling materials as an example, using them as the OSR layer. Raman et al. [82] designed a cooler consisting of seven-layer $SiO_2/HfO_2$ deposited on a silver coated silicon, achieving a solar absorptance of 0.03 and an emittance of ~0.6. Zhai et al. [83] reported a scalable-manufactured metamaterial made of glass-polymer hybrid backed with a silver film. Its emittance reaches 0.93 with solar absorptance as low as 0.04. Yang et al. [84] reported a dual-layer structure with record-high solar reflectance of 0.99 and a high emittance about 0.9. The properties of these materials are marked in **Fig. 6** and the resultant $\Delta\varepsilon_{eff}$ can be immediately read, being 0.38, 0.64 and 0.68, respectively. Therefore, our proposal provides daytime radiative cooling techniques with a new application area, that is, using them in our NFRA smart skin for spacecraft. Overall, one can design or choose $\alpha_S$ and $\varepsilon$ of the OSR, by referring to such results as given by **Fig. 6** in the temperature range of interest, to achieve the required $\Delta\varepsilon_{eff}$.

# 5 Conclusions

We introduce a near-field radiation assisted (NFRA) smart skin which can tune the heat dissipation flux accurately as well as in a large range for spacecraft thermal control. By resorting to MIS structure, the near-field radiative heat flux between it and matching layer can be tuned through the applied voltage. The proposed scheme shows high performance, especially in the in-dark case, with $\Delta\varepsilon_{\text{eff}}$ attaining 0.71 at 270 K while larger than 0.65 at all temperatures of interest. Even in the in-sun case, $\Delta\varepsilon_{\text{eff}}$ would not fall below 0.35 for $T_{\text{H}} \geq 270$ K. Such a significant tuning effect is attributed to the dramatic changes in the near-field heat flux spectrum under different applied voltages. We also demonstrate that the presented smart skin can maintain or control the surface temperature under various thermal conditions. Moreover, by increasing the thermal emittance and decreasing the solar absorptance of the outmost OSR, the tuning range of the effective emittance can be further amplified. It is worth mentioning that, this work focuses on the proposed new principle for NFRA smart skin design and has not conducted many optimizations about materials and structures. Therefore, there is still plenty of room for further improving of the performance of NFRA smart skin for spacecraft thermal control.

## Acknowledgements

The support of this work by the National Natural Science Foundation of China (No. 51976045) is gratefully acknowledged.

## Appendix

**Supplementary Information**

[7] D. Douglas, T. Michalek, T. Swanson, Design of the Thermal Control System for the Space Technology 5 Microsatellite, in: 31st International Conference On Environmental Systems, 2001, pp. 2001-2001-2214.

[8] R. Siegel, J.R. Howell, Thermal Radiation Heat Transfer, Taylor & Francis, New York, 2002.

[9] W. Biter, S. Oh, S. Hess, Electrostatic switched radiator for space based thermal control, AIP Conference Proceedings, 608 (2002) 73-80.

[10] W. Biter, S. Oh, Performance Results of the ESR from the Space Technology 5 Satellites, AIP Conference Proceedings, 880 (2007) 59-65.

[11] S.W. Janson, M.A. Beasley, A.K. Henning, S.L. Firebaugh, R.L. Edwards, A.C. Keeney, R. Osiander, MEMS thermal switch for spacecraft thermal control, in: MEMS/MOEMS Components and Their Applications, Proceedings of the SPIE, Volume 5344, 2004, pp. 98-105.

[12] J. Currano, S. Moghaddam, J. Lawler, J. Kim, Performance Analysis of an Electrostatic Switched Radiator Using Heat-Flux-Based Emissivity Measurement, Journal of Thermophysics and Heat Transfer, 22(3) (2008) 360-365.

[13] A. Hendaoui, N. Émond, M. Chaker, É. Haddad, Highly tunable-emittance radiator based on semiconductor-metal transition of VO2 thin films, Applied Physics Letters, 102(6) (2013) 061107.

[14] A. Hendaoui, N. Émond, S. Dorval, M. Chaker, E. Haddad, VO2-based smart coatings with improved emittance-switching properties for an energy-efficient near room-temperature thermal control of spacecrafts, Solar Energy Materials and Solar Cells, 117 (2013) 494-498.

[15] H. Kim, K. Cheung, R.C.Y. Auyeung, D.E. Wilson, K.M. Charipar, A. Pique, N.A. Charipar, VO2-based switchable radiator for spacecraft thermal control, Sci Rep, 9(1) (2019) 11329.

[16] Y. Shimakawa, T. Yoshitake, Y. Kubo, T. Machida, K. Shinagawa, A. Okamoto, Y. Nakamura, A. Ochi, S. Tachikawa, A. Ohnishi, A variable-emittance radiator based on a metal–insulator transition of (La,Sr)MnO3 thin films, Applied Physics Letters, 80(25) (2002) 4864-4866.

[17] G. Tang, Y. Yu, Y. Cao, W. Chen, The thermochromic properties of La1−xSrxMnO3 compounds, Solar Energy Materials and Solar Cells, 92(10) (2008) 1298-1301.

[18] G. Tang, Y. Yu, W. Chen, Y. Cao, Thermochromic properties of manganese oxides La1−xAxMnO3 (A=Ca, Ba), Materials Letters, 62(17-18) (2008) 2914-2916.

[19] C.H. Wu, J.W. Qiu, M. Xu, J.B. Wang, H.P. Zuo, B.S. Zhang, L. Li, Y.Z. Zhao, Optimization of Thermal Emittance Tuneability of La (Sr, Ca)MnO$_3$ Thin-Film Materials in 173-373 K, Key Engineering Materials, 575-576 (2013) 297-301.

[20] S. Tachikawa, A. Ohnishi, Y. Shimakawa, A. Ochi, A. Okamoto, Y. Nakamura, Development of a Variable Emittance Radiator Based on a Perovskite Manganese Oxide, Journal of Thermophysics and Heat Transfer, 17(2) (2003) 264-268.

[21] N. Kislov, Electrochromic Variable Emittance Devices on Silicon Wafer for Spacecraft Thermal Control, AIP Conference Proceedings, 699 (2004) 112-118.

[22] H. Demiryont, D. Moorehead, Electrochromic emissivity modulator for spacecraft thermal management, Solar Energy Materials and Solar Cells, 93(12) (2009) 2075-2078.

[23] P. Chandrasekhar, B.J. Zay, T. McQueeney, A. Scara, D. Ross, G.C. Birur, S. Haapanen, L. Kauder, T. Swanson, D. Douglas, Conducting Polymer (CP) infrared electrochromics in spacecraft thermal control and military applications, Synthetic Metals, 135-136 (2003) 23-24.

[24] P. Chandrasekhar, B.J. Zay, D. Lawrence, E. Caldwell, R. Sheth, R. Stephan, J. Cornwell, Variable-emittance infrared electrochromic skins combining unique conducting polymers, ionic liquid electrolytes, microporous polymer membranes, and semiconductor/polymer coatings, for spacecraft thermal control, Journal of Applied Polymer Scienc, 131(19) (2014) 40850.

[25] Y. Tian, X. Zhang, S. Dou, L. Zhang, H. Zhang, H. Lv, L. Wang, J. Zhao, Y. Li, A comprehensive study of electrochromic device with variable infrared emissivity based on polyaniline conducting polymer, Solar Energy Materials and Solar Cells, 170 (2017) 120-126.

[26] N. Athanasopoulos, J. Farmasonis, N.J. Siakavellas, Preliminary design and comparative study of thermal control in a nanosatellite through smart variable emissivity surfaces, Proceedings of the Institution of Mechanical Engineers, Part